\documentclass[sn-mathphys-num]{sn-jnl}

\usepackage{graphicx}%
\usepackage{multirow}%
\usepackage{amsmath,amssymb,amsfonts}%
\usepackage{amsthm}%
\usepackage{mathrsfs}%
\usepackage[title]{appendix}%
\usepackage{xcolor}%
\usepackage{textcomp}%
\usepackage{manyfoot}%
\usepackage{booktabs}%
\usepackage{algorithm}%
\usepackage{algorithmicx}%
\usepackage{algpseudocode}%
\usepackage{listings}%

\usepackage{bm}

\makeatletter
\@ifundefined{showcaptionsetup}{}{%
 \PassOptionsToPackage{caption=false}{subfig}}
\usepackage{subfig}
\makeatother

\begin{document}

\title[Assessing the 3D resolution of refocused correlation plenoptic images using a general-purpose image quality estimator]{Assessing the 3D resolution of refocused correlation plenoptic images using a general-purpose image quality estimator}

\author*[1,2]{\fnm{Gianlorenzo} \sur{Massaro}}\email{gianlorenzo.massaro@uniba.it}

\affil*[1]{\orgdiv{Dipartimento Interateneo di Fisica}, \orgname{Università degli Studi di Bari}, \orgaddress{\street{Via Giovanni Amendola, 173}, \city{Bari}, \postcode{70125}, \country{Italy}}}

\affil[2]{\orgname{Istituto Nazionale di Fisica Nucleare, Sezione di Bari}, \orgaddress{\street{Via Giovanni Amendola, 173}, \city{Bari}, \postcode{70125}, \country{Italy}}}

\abstract{Correlation plenoptic imaging (CPI) is emerging as a promising approach
to light-field imaging (LFI), a technique enabling simultaneous measurement
of light intensity distribution and propagation direction from a scene.
LFI allows single-shot 3D sampling, offering fast 3D reconstruction
for a wide range of applications. However, the array of micro-lenses
typically used in LFI to obtain 3D information limits image resolution,
which rapidly declines with enhanced volumetric reconstruction capabilities.
CPI addresses this limitation by decoupling light-field information
measurement using two photodetectors with spatial resolution, eliminating
the need for micro-lenses. 3D information is encoded in a four-dimensional
correlation function, which is decoded in post-processing to reconstruct
images without the resolution loss seen in conventional LFI. This
paper evaluates the tomographic performance of CPI, demonstrating
that the refocusing reconstruction method provides axial sectioning
capabilities comparable to conventional imaging systems. A general-purpose
analytical approach based on image fidelity is proposed to quantitatively
study axial and lateral resolution. This analysis fully characterizes
the volumetric resolution of any CPI architecture, offering a comprehensive
evaluation of its imaging performance.
}


\maketitle

\section{Introduction}

Correlations of light have long been a subject of study due to their
potential to enhance the capabilities of traditional measurement techniques
\citep{Maccone2011,Zavatta2006,Iskhakov2011special,avella2016absolute,Agliati2005,allevi2012measurung,pittman1995optical,pittman1996two}.
In both classical and quantum contexts, the exploration of correlations
has led to significant advancements, particularly in imaging technologies
\citep{Agafonov2009,aspden2013epr,TammaOE,bai2007ghost,bina2013backscattering,brida2011systematic,bromberg2009ghost,chan2009twocolor,d2005resolution,erkmen2010ghost,ferri2005high,TammaSR,ferri2010differential,shih2016physics,xu20181000,paniate2024lfgi}.
In the quantum domain, the unique properties of entanglement and correlation
have been harnessed to surpass the sensitivity limits of conventional
imaging methods \citep{moreau2019imaging}. This has enabled breakthroughs,
such as sub-shot-noise microscopy \citep{brida2010experimental,samantaray2017realization},
providing unprecedented precision in imaging amplitude and phase samples
\citep{Ortolano:2023aa}. Interestingly, correlation properties similar
to those obtained with quantum states of light can also be observed
in classical systems. This convergence of quantum and classical approaches
has revealed that many protocols initially designed for quantum applications
can be effectively adapted to classical contexts \citep{Agafonov2009,allevi2017super,gatti2004ghost,osullivan2010comparison,PhysRevLett.94.063601}.
Consequently, the study of correlations continues to bridge the gap
between quantum and classical imaging, offering versatile solutions
that transcend the traditional boundaries of these domains. Correlation
plenoptic imaging (CPI) \citep{dangelo2016correlation,massaro2021lightfield,massaro2023correlated,abbattista2021towards,Scattarella:2023ab,petrelli2023compressive,pepe2017diffraction}
is emerging as a promising correlation-based approach to light-field
imaging (LFI). LFI is a technique that allows for the concurrent measurement
of both light intensity distribution and propagation direction of
light rays from a three-dimensional scene of interest \citep{adelson1992single}.
The extensive amount of information collected by a light-field device
enables single-shot 3D sampling, a task that would require multiple
acquisitions across various planes with a standard camera \citep{pawley2006handbook,ihrke2016principles}.
This scanning-free characteristic makes light-field imaging one of
the fastest methods for 3D reconstruction, with applications spanning
diverse fields such as photography \citep{ng2005light,ng2005fourier,birklbauer2014panorama},
microscopy \citep{broxton2013wave} and real-time imaging of neuronal
activity \citep{prevedel2014simultaneous}. In its typical implementation,
light-field imaging employs an array of micro-lenses positioned between
the sensor and the imaging device (e.g., the camera lens). These lenslets
generate a series of \textquotedblleft sub-images\textquotedblright{}
corresponding to different propagation directions. However, the presence
of the array significantly limits image resolution, preventing it
from reaching the diffraction limit and causing a rapid decline in
resolution as 3D reconstruction capabilities improve \citep{georgiev2006spatio,goldlucke2015plenoptic}.
CPI addresses the main limitation of conventional LFI by decoupling
the measurement of light-field information on two photodetectors endowed
with spatial resolution \citep{dangelo2016correlation}, in a lenslets-free
optical design. In fact, three-dimensional information about the sample
is encoded in the four-dimensional correlation function, obtained
by correlating the instantaneous light intensity impinging on the
sensors and performing statistical averaging. The correlation function
can then be decoded, entirely in post processing, to reconstruct high-resolution
images of the object without the loss of resolution typical of conventional
LFI.

In this paper, we shall evaluate the tomographic performance of CPI,
showing that the reconstruction approach called \emph{refocusing }\citep{massaro2022refocusing}
endows CPI with the same axial sectioning capabilities of a conventional
imaging system, in which fine axial sectioning is obtained by increasing
the size of the optical elements. After a quantitative study of the
axial depth of the reconstructed images, an analytical approach based
on the \emph{image fidelity} \citep{massaro2024coherent} shall be
proposed. Through the fidelity analysis, the axial and lateral resolution
of CPI be studied quantitatively, in a more general and formally sound
approach, compared to the past, and can be applied to evaluate the
imaging performance of any CPI architecture. Such mathematical tool
will thus be used to fully characterize, analytically and through
numerical analysis, the 3D resolution of CPI.

\section{Imagine reconstruction in CPI}

In CPI, light-field information about the sample is collected by measuring
intensity (or photon number) correlations on two detectors with spatial
resolution. The measured correlation function reads
\begin{equation}
\Gamma(\bm{r}_{a},\bm{r}_{b})=\left\langle I_{A}(\bm{r}_{a})I_{B}(\bm{r}_{b})\right\rangle -\kappa\left\langle I_{A}(\bm{r}_{a})\right\rangle \left\langle I_{B}(\bm{r}_{b})\right\rangle ,\label{eq:corrFunctionDefinition}
\end{equation}
where $I_{A,B}$ is the instantaneous light intensity on the detectors
\citep{s22072778}, and $\bm{r}_{a,b}=(x_{a,b},y_{a,b})$ are the
two-dimensional coordinate on the two detectors surface; the symbol
$\left\langle I\right\rangle $ denotes the ensemble average of the
statistical quantity $I$. The constant $\kappa=0,1$ can vary according
to the illumination source of choice for performing CPI: in fact,
when illuminating with entangled photons, light-field information
is collected by evaluating photon number correlations ($\kappa=0$)
\citep{pepe2016correlation,di2018correlation,dilena2020correlation},
whereas, for thermal and pseudo-thermal light, intensity \emph{fluctuations}
should be measured ($\kappa=1$) \citep{abbattista2021towards,dilena2020correlation,GIANNELLA2024129298,massaro2021lightfield}.
Without loss of generality, we shall henceforth assume that CPI is
performed with pseudo-thermal illumination ($\kappa=1$), and shall
also neglect the four-dimensional dependence of the correlation function,
to limit ourselves only to the $x$-component of the detectors coordinate.\\
Despite CPI can be implemented in many possible variations, from a
fundamental point of view, the mathematical description of the correlation
function can be explained by means of a second-order response function
$\Phi^{\prime}$: in fact, regardless of the particular architecture
\begin{equation}
\Gamma(x_{a},x_{b})=\left|\int A(x_{s})\int\left[A^{*}(x_{s}^{\prime})\right]^{m}\Phi^{\prime}(x_{s},x_{s}^{\prime},x_{a},x_{b})\,dx_{s}^{\prime}\,dx_{s}\right|^{2}.\label{eq:secondOrderResponse}
\end{equation}
Eq. (\ref{eq:secondOrderResponse}) establishes the relationship between
the electric field at the detectors coordinates $x_{a,b}$, and the
field at the sample coordinates $x_{s}$\footnote{Eq. (4) in Ref. \citep{s22072778}};
the function $A(x_{s})$ represents the complex electric field transmittance
of a flat sample, and the coefficient $m$ can either be $0$, if
only one of the two detectors collects light from the object, or $1$
if light from the sample illuminates both sensors \citep{s22072778}.
In the most general case, the optical response of CPI is thus strongly
non-linear with respect to the input function $A$, both because $A$
is involved twice due to a second-order response function, and because
of the square module taken after the integrals. In many cases of interest,
however, such degree of complexity is not required to fully predict
the optical performance of CPI. For instance, when only one detector
collects light from the sample ($m=0$), the system is described by
a \emph{complex} first-order response function (or complex point spread
function, PSF) $\Phi(x_{s},x_{a},x_{b})=\int\Phi(x_{s},x_{s}^{\prime},x_{a},x_{b})$.
On the other hand, for CPI schemes characterized by $m=1$, a second-order
response function results from effects of partial coherence on the
sample surface, namely, from non-negligible coherence area on the
object. Such effects, however, can either be neglected, as is the
case\emph{ }when the sample itself is the source of thermal light,
or should be avoided through careful optical design, \emph{e.g.} when
working with transmissive samples. By neglecting partial coherence
on the source, the response function of the system becomes $\Phi^{\prime}(x_{s},x_{s}^{\prime},x_{a},x_{b})\sim\delta(x_{s}-x_{s}^{\prime})\Phi(x_{s},x_{a},x_{b})$,
so that Eq. (\ref{eq:secondOrderResponse}) can be written as
\begin{equation}
\Gamma(x_{a},x_{b})=\left|\int A(x_{s})\left[A^{*}(x_{s})\right]^{m}\Phi(x_{s},x_{a},x_{b})\,dx_{s}\right|^{2}.\label{eq:responseFunction}
\end{equation}
From Eq. (\ref{eq:responseFunction}), one immediately recognizes
that the only difference between architectures characterized by $m=0$
and $m=1$ is that the former are sensitive to the phase content of
the complex function $A$, whereas the latter are only sensitive to
the intensity profile of the object $\left|A\right|^{2}$. This distinction
does not have any influence on the optical performance of the technique,
so we shall limit our discussion to phase-insensitive architectures,
characterized by negligible coherence area on the sample surface.

To some extent, some information about the sample can be inferred
already from the correlation function itself \citep{s22072778}; however,
in order to gain axial localization of the sample features and dramatically
improve the signal to noise ratio (SNR)\citep{massaro2022comparative,scala2019signal},
the $\Gamma$ function must be \emph{refocused }to obtain a sharp
image of the sample. As outlined in Refs. \citep{massaro2022refocusing,s22072778,di2018correlation},
the refocusing procedure is based on a reasoning that is based entirely
on geometrical optics arguments, \emph{i.e.} on ray-tracing. Through
refocusing, a specific object coordinate $(x_{s},z)$ is reconstructed
by summing together all the correlated optical paths leading to the
two detectors, crossing the object plane at axial coordinate $z$
and transverse coordinate $x_{s}$. In CPI, the geometrical locus
of detector points corresponding to a single coordinate $(x_{s},z)$
is a straight line in the $(x_{a},x_{b})$ plane, so that the reconstruction
is obtained through the line integral
\begin{equation}
\Sigma_{z}(x_{s})=\int_{\gamma(x_{s},z)}\Gamma\,d\ell,\label{eq:refocusDefinition}
\end{equation}
where the integration path $\gamma(x_{s},z)$ has equation
\begin{equation}
\gamma(x_{s},z):\alpha(z)\,x_{a}+\beta(z)\,x_{b}=x_{s}\label{eq:integration_path}
\end{equation}
in the $(x_{a},x_{b})$ plane. As specified in Eq. (\ref{eq:integration_path}),
the axial plane that is reconstructed is determined by the coefficients
$\alpha(z)$ and $\beta(z)$, which, according to the optical design
of the experiment, can be associated to the optical distances and
and experimental parameters.

\section{Axial sectioning}

Because of the geometry of the integration path reconstructing a sample
coordinate $x_{s}$, Eq. (\ref{eq:refocusDefinition}) can be recognized
as a Radon transform of the correlation function at an angle $\theta(z)=-\arctan\left[\alpha(z)/\beta(z)\right]$
so that, in the formalism conventionally used for techniques based
on tomographic reconstruction through Radon transformation \citep{5484190},
\begin{equation}
\Sigma_{z}(x_{s})=R\Gamma(\theta,x_{s})=\int_{-\infty}^{+\infty}\Gamma\left(x^{\prime}\,\sin\theta+x_{s}\,\cos\theta,\ -x^{\prime}\,\cos\theta+x_{s}\,\sin\theta\right)dx^{\prime}.\label{eq:radon}
\end{equation}
Such an analogy allows us to promptly infer that, as with any tomographic
technique based on back-projection reconstruction, the ``depth''
of the reconstructed images can be expected to depend on the number
of directions in the Fourier domain available for the reconstruction
\citep{levoy1992volume,ng2005fourier}, or, equivalently, on the length
of the integration path $\gamma(x_{s},z)$ in Eq. (\ref{eq:refocusDefinition}).
In other words, the axial sectioning of CPI can be associated to the
extension of the support of the correlation function which, as demonstrated
both theoretically \citep{s22072778,massaro2022refocusing} and experimentally
\citep{massaro2023correlated}, depends on the size of the optics,
namely, on the numerical aperture (NA).

\begin{figure}
\includegraphics[width=0.5\textwidth]{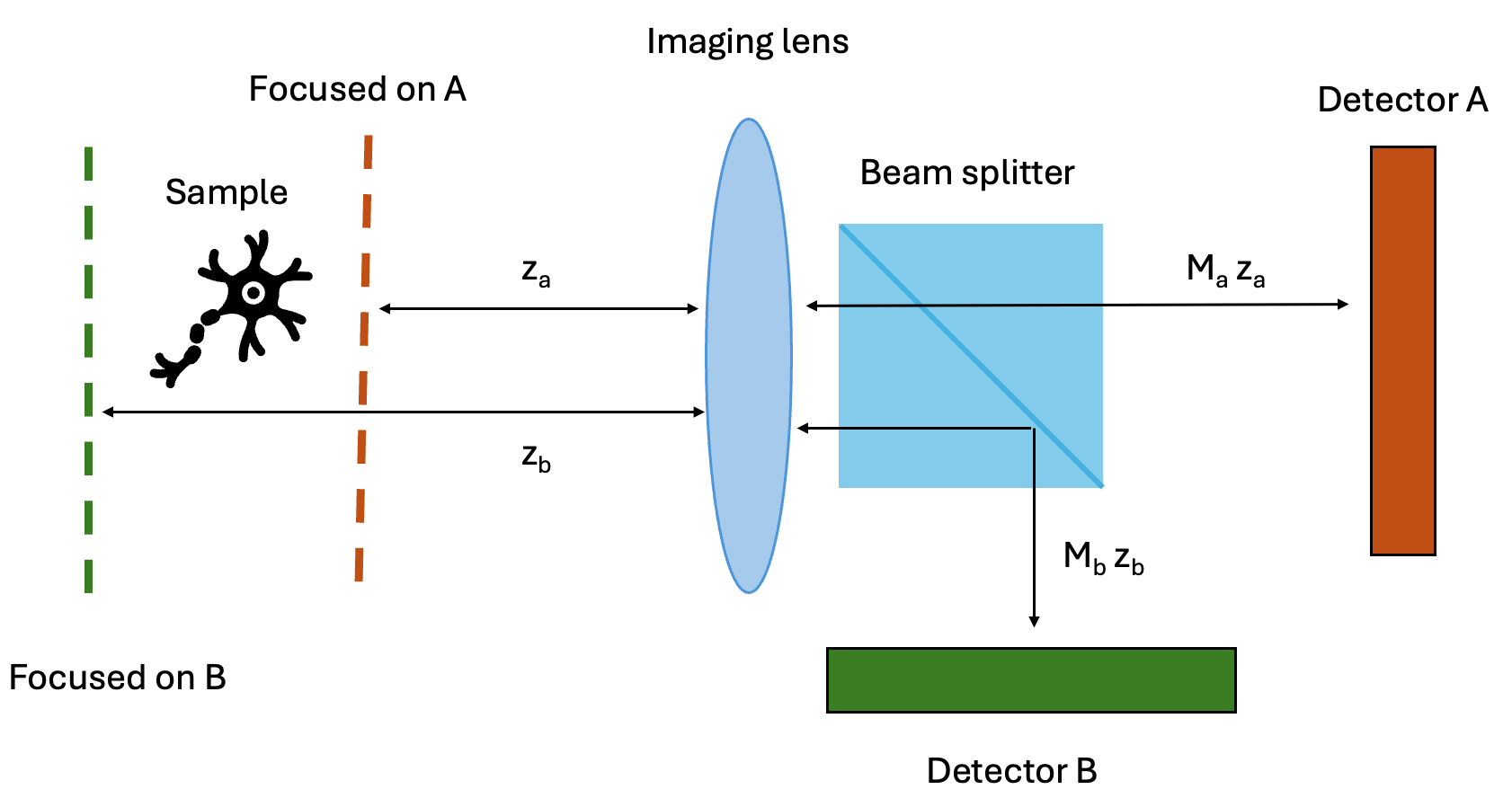}

\caption{Schematics of a single-lens CPI scheme\label{fig:setup}}
\end{figure}

For definiteness, all the results presented in the paper are obtained
on a CPI architecture based on a single-lens design \citep{dilena2020correlation},
schematically represented in Fig. \ref{fig:setup}. In this architecture,
the two detectors $A$ and $B$ are optically conjugated, by means
of an imaging lens, to two planes in the region of space occupied
by the sample. We shall indicate with $z_{A}$ ($z_{B}$) the optical
distance between the object plane imaged on detector $A$ ($B$) and
the lens, so that the plane is imaged with magnification $M_{A}$
($M_{B}$), as defined by the optical distances involved and the focal
length $f$ of the lens. After the lens, light coming from the sample
is separated into to two optical paths leading towards the detectors
by means of a beam-splitter.

The tomographic capabilities of CPI are better understood through
direct comparison with a conventional imaging system, based on a single-lens
design and on intensity measurement (for instance, if only measuring
the intensity $I_{A}$ impinging on detector A). In such systems,
the axial scanning of the 3D space surrounding the sample is performed
mechanically, by varying the relative distance between the detector
and lens, so that a stack of different planes at focus is available
after the measurement. The quality of the axial scanning depends on
how effectively the imaging system suppresses, at any given focusing
position, spurious contributions from unfocused planes, namely, on
the thickness of the depth of field (DOF). Suppression of background
planes in conventional imaging is knowingly due to the circle of confusion
(COC) \citep{Stokseth:69}, a mechanism that is entirely explained
in terms of ray optics, which makes so that point-like objects in
the background produce a circle-shape projection onto the plane on
focus. Such effect is increasingly more relevant as the NA of the
system increases, so that high-NA designs not only beneficial because
of the high resolution they provide at focus, but also because of
their shallow DOF \citep{microscopy_chap6}. We shall now demonstrate
that CPI benefits from the same high-NA design, as far as axial sectioning
is concerned.
\begin{figure}
\subfloat[\label{fig:COC_corrFun}]{\includegraphics[width=0.5\textwidth]{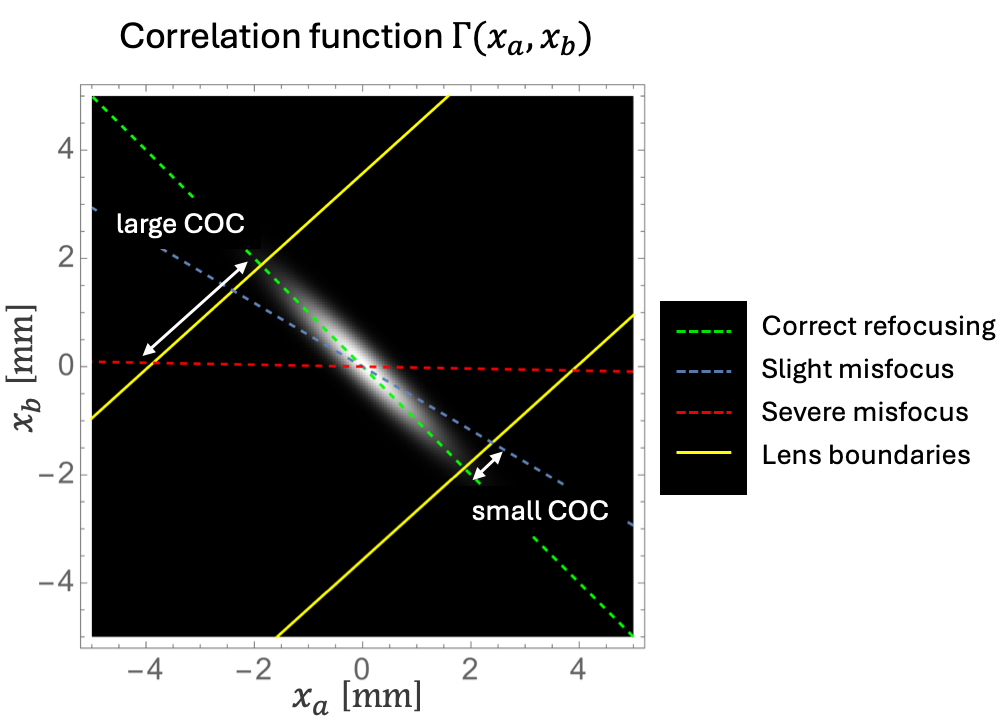}

}\subfloat[\label{fig:COC_refoc}]{\includegraphics[width=0.45\textwidth]{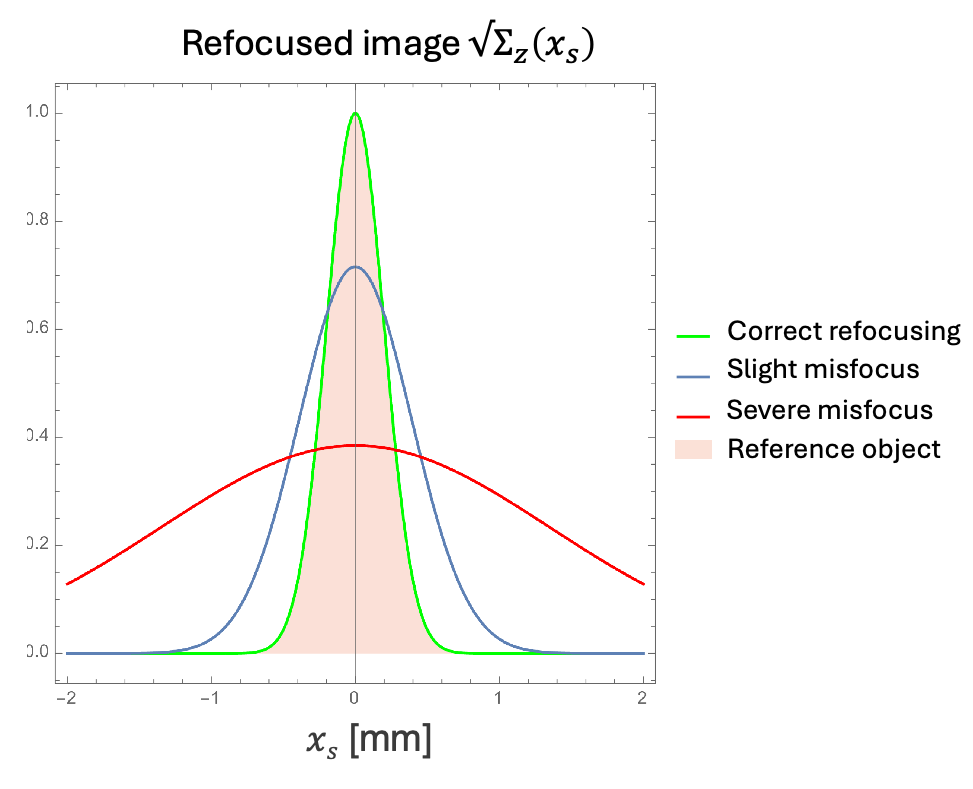}

}

\caption{\emph{Panel} a.: measured correlation function in the case of a Gaussian-slit
object, placed on the optical axis at $z_{s}=(z_{a}+z_{b})/2$;\emph{
Panel} b.: comparison between a proper refocusing $\sqrt{\Sigma_{z}}$
with $z=z_{s}$ and an incorrect refocusing at $z\protect\neq z_{s}$}

\end{figure}
 Fig. \ref{fig:COC_corrFun} reports an example of a measured correlation
function $\Gamma(x_{a},x_{b})$, obtained by simulating the optical
setup in Fig. \ref{fig:setup}. To obtain the correlation function,
we set the focal length of the imaging lens to $f=75\,$ mm; detectors
A and B image their respective conjugate planes with magnifications
$M_{a}=1.1$ and $M_{b}=0.9$, thus resulting in $z_{a}\simeq143$
mm and $z_{b}\simeq158$ mm. For ease of calculation, we have assumed
the lens pupil to have a Gaussian apodization, with a width of $\sigma_{l}=25$
mm. The plot is obtained by assuming that the imaged object is a Gaussian
transmissive slit having intensity profile
\begin{equation}
\left|A(x_{s})\right|^{2}=\exp\left[-\frac{x_{s}^{2}}{2\sigma_{s}^{2}}\right],\label{eq:gaussianObject}
\end{equation}
with $\sigma_{s}=200\,\mu$m. For the case shown in Fig. \ref{fig:COC_corrFun},
the object is placed on the optical axis at a distance $z_{s}=(z_{a}+z_{b})/2$
from the lens, so that neither A nor B yield a focused image of the
sample. For this CPI architecture, the object coordinate $x_{s}$
of a sample placed at a generic axial distance $z$ is reconstructed
by integrating along the line
\begin{equation}
\gamma(x_{s},z):x_{s}=-\frac{z-z_{b}}{z_{a}-z_{b}}\,\frac{x_{a}}{M_{a}}+\frac{z-z_{a}}{z_{a}-z_{b}}\,\frac{x_{b}}{M_{b}}.
\end{equation}
Therefore, an object placed mid-way between $z_{a}$ and $z_{b}$
is correctly reconstructed via a Radon transformation at an angle
$\theta(z=z_{s})=-\pi/4$, as is evident from the picture (the dashed
green line is parallel to the object features). As is the case for
standard imaging, however, one does not typically know \emph{a priori
}the correct axial position of the sample. Hence, a complete $z$-scan
of the 3D space surrounding the sample is usually needed. This entails
calculating Radon transformations at many different angles $\theta(z)\neq\theta(z_{s})$
which can be expected to result in a less sharp image than perfect
refocus $\theta(z_{s})$, and also to a suppression of the intensity
peak, as is the case for out-of-focus imaging in a conventional scenario.
Rather intuitively, this effect must be the analogous of the COC,
since the blurring of misfocused plane can easily be envisioned to
depend on how extended the integration paths are (as defined by the
NA), and on how misfocused the integration angle is with respect to
the correct one. Such picture can be graphically verified from Fig.
\ref{fig:COC_corrFun}, where we reported the lens boundaries (yellow
lines, corresponding to the projection of the lens coordinates at
$\pm3/2\,\sigma_{l}$ onto the $(x_{a},x_{b})$ plane), and two integration
angles corresponding to $z=152\,$ mm $\neq z_{s}$ (dashed blue line),
and $z=z_{b}\neq z_{s}$ (dashed red line). In fact, the extent of
blurring can clearly be expected to be directly proportional to both
the difference of the lines slope, with respect to the green line,
and to how large the NA of the lens is (distance between the yellow
lines). This is verified in Fig. \ref{fig:COC_refoc}, demonstrating
that the best reconstruction (green line) happens when integrating
at $\theta(z_{z})$ and results in an image indistinguishable from
the reference object (solid salmon area). For the other integration
directions (solid blue and red lines), slight and severe blurring
occur, respectively, so that the final reconstruction results in a
blurred and fainter image.

\subsection{Quantitative analysis of the axial sectioning}

From a qualitative point of view, the results shown in Figs. \ref{fig:COC_corrFun}
and \ref{fig:COC_refoc} hint that the tomographic capabilities of
the refocusing algorithm are related to the COC. This can also be
proved quantitatively, by studying how faithfully can CPI reconstruct
the image of an object placed at a given axial coordinate. In order
to do so, we shall use a newly introduced tool for evaluating image
quality, named the \emph{image fidelity} \citep{massaro2024coherent}.
If an imaging system produces an image $I(x_{s})$ of a sample with
intensity transmittance $\left|A(x_{s})\right|^{2}$, the fidelity
$F_{A}[I]$ is defined as
\begin{equation}
F_{A}[I]=\int\sqrt{I(x)\left|A(x)\right|^{2}}\,dx;\label{eq:fidelity_definition}
\end{equation}
where the quantities involved must be properly normalized ($\int I(x_{s})\,dx_{s}=1$
and $\int\left|A(x_{s})\right|^{2}\,dx_{s}$). $F_{A}[I]$ saturates
to unity for perfectly accurate imaging systems, namely $I(x)=\left|A(x)\right|^{2}$,
and approaches $0$ for very unfaithful imaging. The analysis in terms
of fidelity enables us to obtain \emph{fidelity curves }that fully
characterize the optical performance of the technique in terms of
the experimental parameters, and replace the concept of \emph{resolution}
in imaging modalities where such concept cannot be as clearly defined
as in standard imaging \citep{massaro2024coherent}.

To fully characterize the analogies with the COC of a misfocused imaging
system, we shall compare the fidelity of a CPI refocusing $z$-scan
directly with the fidelity of the mechanical $z$-scan of a standard
imaging system, having the same optical design as a single arm of
our CPI scheme. If an object of size $\sigma_{s}$ is fixed at a distance
$z_{s}$ from the lens, an axial scan is obtained by moving the detector-to-lens
distance so as to change the $z$ in focus; in this context, the resulting
image depends on both parameters $z_{s}$ and $z$. Overall, the image
resulting from an intensity measurement depends on three parameters,
and we shall indicated as it as $I_{\text{std}}(x;\,z,z_{s},\sigma_{s})$,
where $x$ is the image coordinate. If an object is placed at $z_{s}$,
also the image reconstructed through CPI refocus $\Sigma_{z}(x;\,x_{s},\sigma_{s})$
depends on the same three parameters, with the difference that $z$-scanning
is obtained through Radon transformation and not by mechanical movement
of the detector. For our analysis, both $z_{a}$ and $z_{b}$ will
be considered to be fixed. The fidelity of a standard and refocused
image are evaluated, respectively, as
\begin{align}
F_{A}[I_{\text{\text{std}}}](z,z_{s},\sigma_{s}) & =\int\sqrt{I_{\text{std}}(-M(z)\,x_{s},\,z,z_{s},\sigma_{s})\,\left|A(x_{s},\,\sigma_{s})\right|^{2}}dx_{s}\label{eq:fidelity_std}\\
F_{A}[\Sigma_{\text{z}}](z,z_{s},\sigma_{s}) & =\int\sqrt{\sqrt{\Sigma_{\text{z}}}(x_{s};\,z_{s},\sigma_{s})\,\left|A(x_{s},\,\sigma_{s})\right|^{2}}dx_{s}.\label{eq:fidelity_CPI}
\end{align}
The two definitions are adapted so as to compare the results as fairly
as possible. In fact, the optical magnification of the imaging system
$M(z)$ has been introduced in Eq. (\ref{eq:fidelity_std}) to scale
the image to its original size on the object side of the lens; such
operation is not necessary in CPI, since it is included in the refocusing
procedure \citep{massaro2022refocusing}. Since the $z$-scanning
is obtained by changing the detector position, each plane in focused
is image with a different magnification, originating the dependence
of $M$ on $z$. The additional square root on $\Sigma_{z}$ has been
introduced in Eq. (\ref{eq:fidelity_CPI}) to compensate for the known
fact that, even in perfect imaging conditions, CPI returns the \emph{squared}
intensity profile of the sample when both detectors see light from
the object (see Eq. (\ref{eq:secondOrderResponse}) when $m=1$).
\begin{figure}
\includegraphics[width=0.8\textwidth]{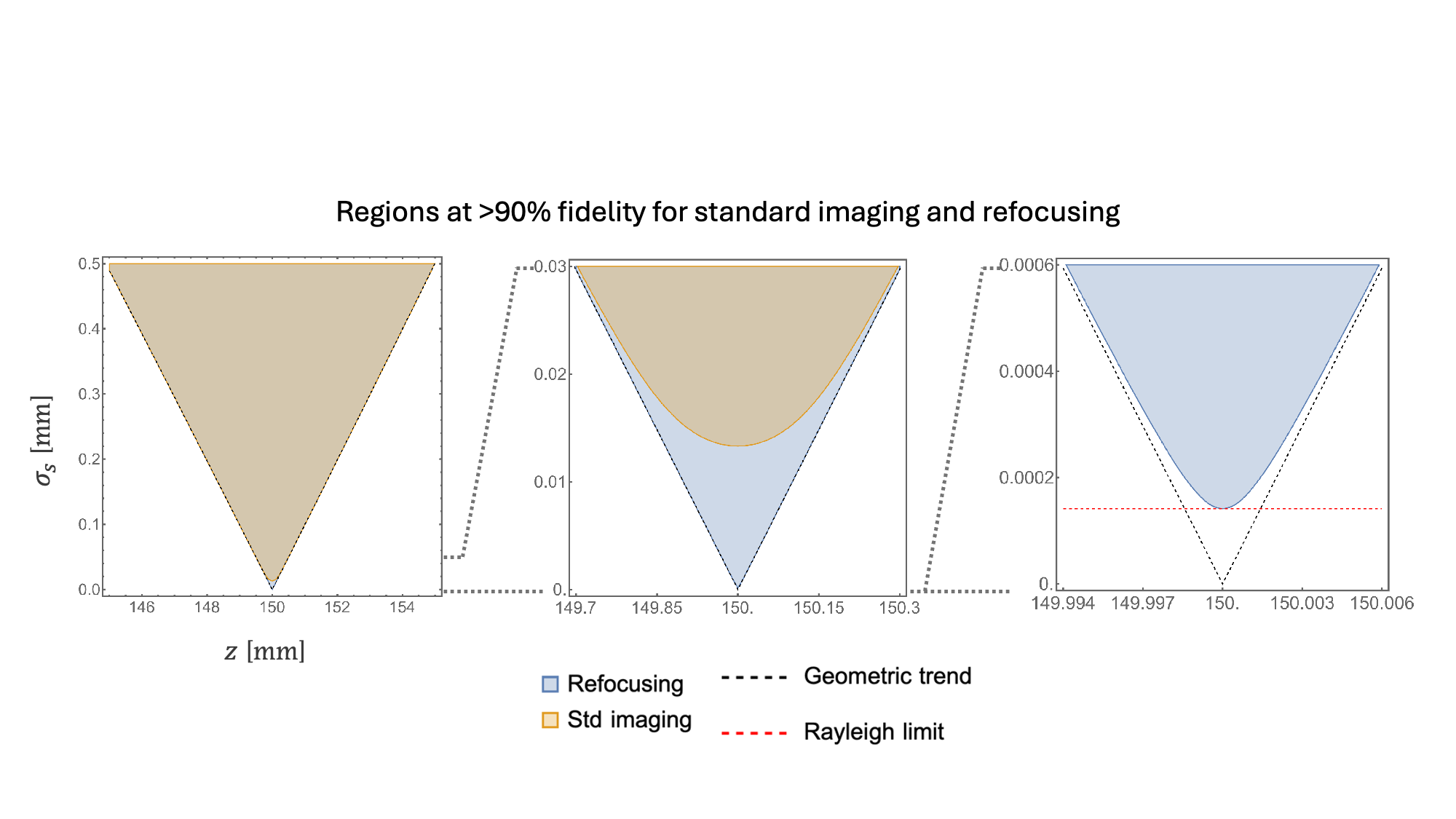}

\caption{Curves characterized by image fidelity larger than $90\%$ in the
$(z,\sigma_{s})$ space\label{fig:axial_fidelity}}

\end{figure}

In Fig. \ref{fig:axial_fidelity}, we report the results of the fidelity
analysis for standard imaging and CPI refocusing. The analysis has
been carried out by considering a Gaussian object of width $\sigma_{s}$,
placed at a fixed distance $z_{s}=(z_{a}+z_{b})/2=150$ mm from the
lens, for both CPI and standard imaging. The experimental parameters
$z_{a}$, $z_{b}$, $f$ and $\sigma_{l}$ have been fixed to the
same values as in Fig. \ref{fig:COC_corrFun}. Since the object is
at a fixed coordinate, the fidelity of Eqs. (\ref{eq:fidelity_std})
and (\ref{eq:fidelity_CPI}) reduces to a two-variable function of
the object size $\sigma_{s}$ and scanning coordinate $z$. The blue
area, corresponding to refocusing, and orange area, corresponding
to standard imaging, identify the region of the $(z,\sigma_{s})$
space in which a sample can be imaged, by the corresponding technique,
with fidelity larger than $90\%$. Hence, at any given object size
$\sigma_{s}$, the difference between the largest and smallest $z$
at which an object is imaged with $90\%$ fidelity can be regarded
as the axial resolution, or DOF, of the technique. As we can see,
for large object sizes and large displacements (leftmost panel), refocusing
shows exactly the same performance as conventional imaging, which
is knowingly limited by the COC. Considering the COC is a ray optics
concept, it is worth investigating the behavior of $F_{A}[\Sigma_{z}]$
in the geometrical optics approximation for refocusing, namely
\begin{equation}
F_{\text{geom}}(z,z_{s},\sigma_{l})=\lim_{\lambda\rightarrow0}F_{A}[\Sigma_{z}](z,z_{s},\sigma_{l}).
\end{equation}
The implicit equation $F_{\text{geom}}(z,z_{s},\sigma_{l})=c$, identifying
the curve at fidelity $c$ in the $(z,z_{s},\sigma_{l})$ space, can
be easily inverted to obtain the minimum size of the object that can
faithfully be reconstructed, as a function $z$ and $z_{s}$
\begin{equation}
\sigma_{\text{geom}}(z,z_{s})\simeq\left|z-z_{s}\right|\frac{\sigma_{l}}{z_{s}}\,f(c),\label{eq:geom_fidelity}
\end{equation}
where $f(c)$ is a function of the threshold value $c$, arbitrarily
chosen to discriminate between faithful and unfaithful images. As
predicted, the expression of $\sigma_{geom}$ is exactly the same
as the COC in conventional imaging, being directly proportional to
both the \emph{effective} numerical aperture $\sigma_{l}/z_{s}$,
and to the relative displacement from the plane of sharpest reconstruction.
As can be seen from Fig. \ref{fig:axial_fidelity}, where Eq. (\ref{eq:geom_fidelity})
is reported as a dashed black line, when the object size is far from
the lateral resolution limit of the techniques at the given $z_{s}$,
the optical performance of both standard imaging and refocusing is
determined by ray optics. In this conditions, one obtains that the
thickness of a reconstructed image at any given object size is
\begin{equation}
\text{DOF}_{\Sigma}(\sigma_{S})\propto\frac{\sigma_{s}}{\text{NA}},
\end{equation}
as for a standard imaging system.

\section{Lateral resolution of CPI}

In the middle panel of Fig. \ref{fig:axial_fidelity}, we see that
when the object size approaches the resolution limit of CPI at the
object position $z_{s}$, the optical performance of refocusing detaches
from the geometrical trend, which would incorrectly prescribe an point-like
resolution ($\sigma_{\text{geom}}(z=z_{s},z_{s})=0$). A conventional
$z$-scanning imaging system, however, still behaves according to
geometrical optics at that scale, detaching from it only when the
Rayleigh limit ($\propto\lambda/$NA) is reached (right panel). The
fact that, at any given transverse plane $z_{s}$, the resolution
of CPI refocusing is worse than a conventional imaging system is a
known fact. For a CPI architecture such as the one we are considering,
in fact, only two planes can be knowingly reconstructed at Rayleigh-limited
resolution, namely, the planes at distance $z_{a}$ and $z_{b}$,
which are optically conjugated to the detectors \citep{dilena2020correlation}.

As opposed to the previous section, we shall now analyze the image
quality of a refocused image as the object is moved along the optical
axis, by disregarding the axial depth of the reconstructions. This
corresponds to studying the case in which, for any given object placement
$z_{s}$ one does not consider the whole $z$-scanning $\Sigma_{z}$,
but only the sharpest reconstruction $\Sigma_{z=z_{s}}.$ Through
Eq.(\ref{eq:fidelity_CPI}), we can thus evaluate the performance
of CPI in the plane of best refocusing, namely $F_{A}[\Sigma_{z}](z=z_{s},z_{s},\sigma_{s})$.
For reference, we shall also report the image quality obtained by
detectors $A$ and $B$ without the use of correlations, to show the
performance improvement granted by the correlation measurement. The
fidelity of the images collected on the detectors, separately, can
be evaluated through Eq. (\ref{eq:fidelity_std}), as $F_{A}\left[I_{\text{std}}\right](z=z_{a,b},z_{s},\sigma_{s})$,
at the two fixed experimental values $z_{a}$ and $z_{b}$.
\begin{figure}
\includegraphics[width=0.8\paperwidth]{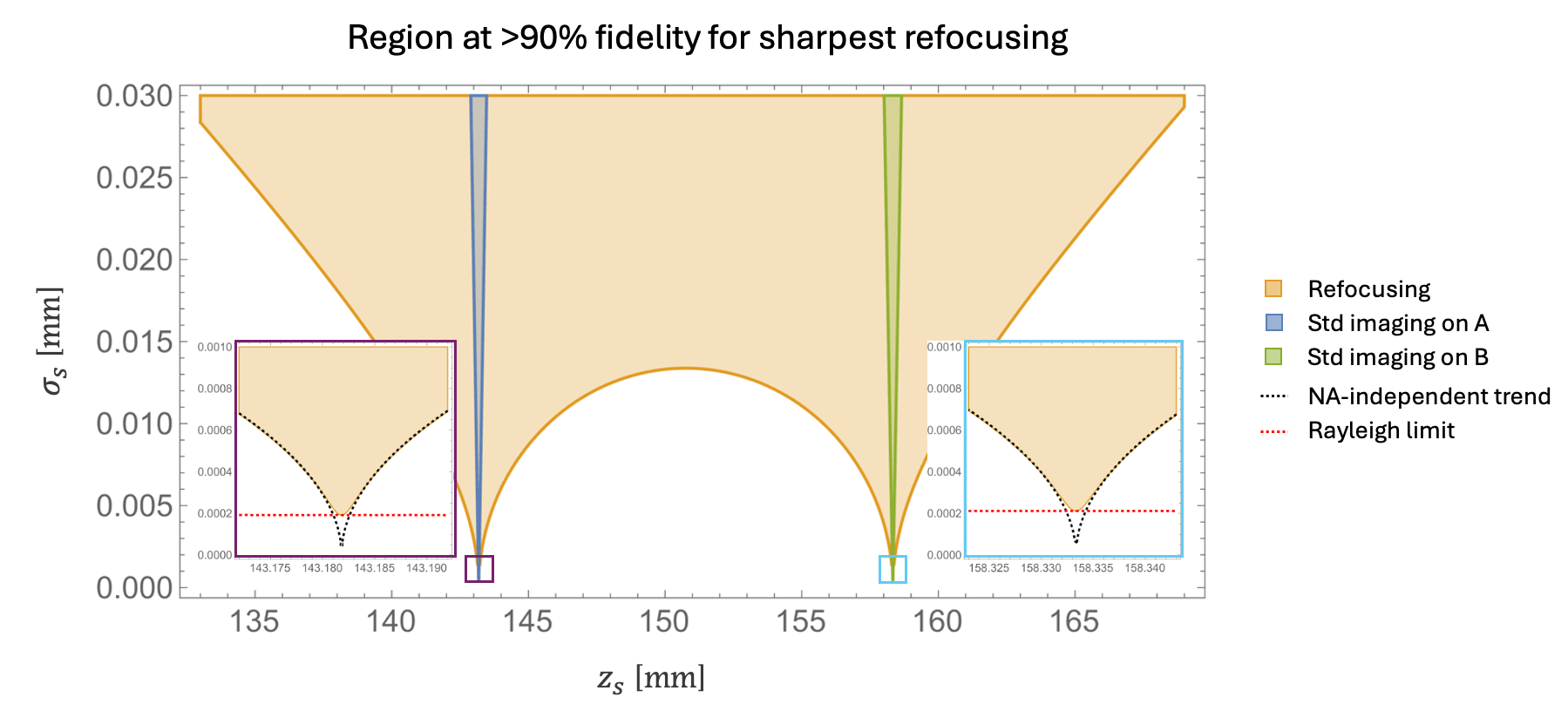}

\caption{Fidelity analysis in the plane of sharpest reconstruction $\Sigma_{z=z_{s}}$\label{fig:fidelity_refocusing}}

\end{figure}
\\
The fidelity analysis in the plane of sharpest reconstruction is reported
in Fig. \ref{fig:fidelity_refocusing}. As already well known, the
DOF of CPI is much larger than the DOF of the two conventional single-lens
systems measuring intensity on A and on B (blue and green areas, respectively).
As in the previous case, we define the DOF as the difference between
the largest and smallest $z_{s}$ that allow for a faithful image
reconstruction at a given object size. We should point out, however,
that the meaning of DOF in this context is different from the previous
section. In fact, the previous DOF was an estimation of the ``depth''
of the reconstruction, whereas now it is the axial range in which
CPI can successfully reconstruct an object of given size $\sigma_{s}$.
The difference can be understood by considering that the previous
analysis was carried out by considering a \emph{fixed object} at coordinate
$z_{s}$, while the analysis reported in Fig. \ref{fig:fidelity_refocusing}
evaluates the best possible refocusing at any object coordinate. In
other terms, if the analysis of Fig. \ref{fig:axial_fidelity} is
useful for evaluating the DOF of a refocused image, Fig. \ref{fig:fidelity_refocusing}
allows one to evaluate the DOF of the technique itself. The DOF of
the reconstruction should then be regarded as the \emph{axial resolution
}of the technique, so that the ratio between the two determines the
number of independent planes in the overall DOF of the technique that
can be \emph{isolated }through refocusing.

The \emph{fidelity curves} of Fig. \ref{fig:fidelity_refocusing}
must be intended as a replacement for the \emph{visibility }or \emph{resolution
curves }of CPI, presented in previous literature \citep{dangelo2016correlation,dilena2020correlation,GIANNELLA2024129298,massaro2021lightfield,massaro2023correlated,di2018correlation}.
For reason that will be clarified shortly, however, the fidelity curves
represent a much more general way of estimating the optical performance
of CPI. The lower boundary to the orange areas of the figure represents
the parametric curve $F_{A}[\Sigma_{z}](z=z_{s},z_{s},\sigma_{s})=0.9$,
namely, the $90\%$-fidelity curve in the $(z_{s},\sigma_{s})$ plane.
Such curve thus establishes a correspondence between the axial coordinate
$z_{s}$ and the minimum object size $\sigma_{\text{min}}(z_{s})$
that can be reconstructed faithfully enough at that position. It is
already known that the functional relationship $\sigma_{\text{min}}(z_{s})$
is for the most part independent of the numerical aperture of the
setup, with the exception of the planes in focus ($z_{s}=z_{a}$ and
$z_{s}=z_{b}$), which are available with Rayleigh-limited resolution.
Such independence of the optical performance of CPI from NA sets the
technique apart from conventional imaging for which, as we detailed
in the previous section, the defocused optical performance is defined
by the NA-proportional COC. One can thus check whether the performance
of CPI can be predicted independently on NA even from an analytical
point of view, by studying the fidelity of an ideal system
\begin{equation}
F_{\infty}(z_{s},\sigma_{s})=\lim_{\text{NA}\rightarrow\infty}F_{A}[\Sigma_{z}](z=z_{s},z_{s},\sigma_{s}).
\end{equation}
Also in this case, the implicit equation of the $c$-fidelity curve
$F_{\infty}(x_{s},\sigma_{l})=c$ can be inverted, so that the fidelity
curve in the infinite-NA case read
\begin{equation}
\sigma_{\infty}(z_{s})\simeq\sqrt{\lambda\left|\frac{1}{z-z_{a}}-\frac{1}{z-z_{b}}\right|^{-1}}\,f^{\prime}(c).\label{eq:resolution_inftyNA}
\end{equation}
The infinite-NA fidelity curve is reported in the two insets in Fig.
\ref{fig:fidelity_refocusing}, showing the details of the CPI fidelity
in very close proximity of the two axial planes in focus. The plot
demonstrates that, apart from a very small region of space in which
the resolution is defined by the NA-dependent Rayleigh limit ($\propto\lambda/$NA),
the optical performance of CPI reproduces with extreme accuracy trend
associated to the infinite-NA regime. The mechanism responsible for
the loss of resolution of CPI outside of the natural DOF of the lens
is thus completely independent of the optics size.

\section{Conclusions}

The surprising independence of the lateral resolution of CPI on the
NA of the imaging system, with the exception of the plane (or planes)
in focus, was already known in previous literature and even proven
experimentally \citep{massaro2023correlated}. The origin of such
property was, however, not clearly understood, mostly because of two
aspects: firstly, the exact mathematical relationship between the
form of the refocused image and the object is not known, and, secondly,
because the image quality of CPI has been so far assessed by using
conventional image quality estimators (\emph{e.g.} ``two-point''
resolution, maxima-minima visibility, modulation transfer function
(MTF) analysis). Such estimators are unfit to correctly evaluate the
performance of CPI: from Eq. (\ref{eq:responseFunction}), it is rather
evident that the technique, even without considering the added degree
of complexity introduced by refocusing, is a \emph{non-linear} imaging
method, in the sense that the input signal $\left|A(x)\right|^{2}$
is not related to the output image through a simple \emph{transfer
function}, as is the case for conventional imaging systems. In standard
imaging, in fact, the only mechanism\footnote{If aberrations are neglected.}
responsible for image quality degradation is blurring or, equivalently,
the fact that the final image is obtained by convolution of the input
with a \emph{positive} PSF. Because of this property, all the merit
figures conventionally used for image quality assessment can be reduced
to the underlying linearity of the imaging formation process, and
can thus be applied to a limited extent to predict the performance
of nonlinear imaging \citep{massaro2024coherent}. In this work, we
have decided to base our analysis of the optical performance of CPI
on the image fidelity, which, being independent of the details of
the image formation process, allows for a direct and \emph{unbiased}
comparison to other techniques. Through the fidelity analysis, we
have demonstrated that the same results and NA-independence of the
lateral resolution in the plane of best refocusing can be obtained
in a mathematically consistent formalism, which has the advantage
of not relying on assumptions on the imaging formation process. Compared
to an assessment through image visibility, typically used in literature
to evaluate the performance of CPI \citep{dilena2020correlation,di2018correlation},
the fidelity has the advantage is based on the \emph{global} features
of the image, and not on the very local nature of maxima and minima.
Furthermore, compared to visibility, the fidelity is always an increasing
function of the object size, when all the other parameters are fixed,
\[
\frac{\partial F_{A}[\Sigma_{z}]}{\partial\sigma_{s}}(z,z_{s},\sigma_{s})>0,
\]
and does not show any unphysical fluctuating behavior \citep{scattarella2022resolution,Scattarella:2023aa,dilena2020correlation,di2018correlation}.
Most importantly, the results that we presented in this work only
for the case of the optical design of Fig. \ref{fig:setup}, have
been obtained also by extending the study to other CPI architectures;
in all cases, the fidelity analysis has confirmed the NA independence
and square-root trend of the lateral resolution of CPI that was already
known in literature. Through the fidelity analysis we have confirmed
that the fidelity curves in the plane of best refocusing are always
given by a square-root law, independently of the architecture
\begin{equation}
\sigma_{\infty}(z_{s})\propto\sqrt{\lambda\,\left|d(z_{s})\right|};
\end{equation}
the ``equivalent distance'' $d(z_{s})$ is, however, a function
of the CPI architecture at hand. For instance, for CPI architectures
based on position-momentum correlations \citep{GIANNELLA2024129298},
$d(z_{s})=z_{s}$, where $z_{s}$ is the distance from focus, so that
the resolution is given by a pure square-root law. For the architecture
discussed here, instead, the infinite-NA trend which correctly reproduces
the fidelity curves 
\begin{equation}
d(z_{s})=\left(\frac{1}{z-z_{a}}-\frac{1}{z-z_{b}}\right)^{-1},
\end{equation}
as can be deduced from Eq. (\ref{eq:resolution_inftyNA}). In all
cases, the performance of CPI outside of the planes in focus can be
mathematically obtained by evaluating the infinite-NA fidelity $F_{\infty}(z_{s},\sigma_{s})$,
and then inverting the implicit equation $F_{\infty}=c$.

Unlike the NA-independent image quality in the plane of sharpest refocusing,
the axial sectioning enabled by refocusing is defined entirely by
the NA, specifically by the same COC defining the imaging depth of
conventional imaging systems. This results in a very interesting fact,
namely, that the lateral and axial resolution are decoupled from each
other. This property enables the designs of optical setups characterized
by high resolution, as defined by the CPI scheme of choice, with the
sectioning capabilities that can be tuned independently by selecting
the appropriate lens size. Such operation can even be performed in
post-processing, by limiting the size of the refocusing integration
path (Eq. (\ref{eq:refocusDefinition})) to simulate a reduced \emph{effective}
NA, with no effect on resolution, for a deeper image reconstruction
and improved computational efficiency. Such versatility sets apart
CPI from conventional, lenslet-based, plenoptic imaging, which suffer
from a strong trade-off between lateral and axial resolution \citep{dansereau2013decoding,georgiev2009high,goldlucke2015plenoptic}.

\backmatter

\bmhead{Acknowledgements}

This article has benefited from the contributions of prof. M. D'Angelo,
who obtained project funding and reviewed the final draft.\smallskip{}

\bmhead{Funding}
The author acknowledges funding from Università degli Studi di Bari
under project ADEQUADE, and from Istituto Nazionale di Fisica Nucleare
under projects Qu3D and QUISS.

Project ADEQUADE has received funding from the European Defence Fund
(EDF) under grant agreement EDF-2021-DIS-RDIS-ADEQUADE (n° 101103417). Project Qu3D
is supported by the Italian Istituto Nazionale di Fisica Nucleare,
the Swiss National Science Foundation (grant 20QT21187716 \textquotedblleft Quantum
3D Imaging at high speed and high resolution\textquotedblright ),
the Greek General Secretariat for Research and Technology, the Czech
Ministry of Education, Youth and Sports, under the QuantERA programme,
which has received funding from the European Union\textquoteright s
Horizon 2020 research and innovation programme.

Funded by the European Union. Views and opinions expressed are however
those of the author(s) only and do not necessarily reflect those of
the European Union or the European Commission. Neither the European
Union nor the granting authority can be held responsible for them.

\bmhead{Conflict of interest}
The author has no competing interests to declare that are relevant to the content of this article.

\bmhead{Data availability}
Not applicable

\bigskip

%
%






\bibliography{Bibliography}

\end{document}